\documentclass[%
 reprint,
 superscriptaddress,
 amsmath,amssymb,
 aps,
 prl,
]{revtex4-1}

\usepackage{graphicx}
\usepackage{dcolumn}
\usepackage{bm}
\usepackage{colortbl}
\usepackage[table]{xcolor}
\usepackage{makecell}
\usepackage[mathscr]{euscript}
\usepackage[table]{xcolor}
\usepackage{multirow}
\usepackage{braket}
\usepackage[pagewise]{lineno}
\usepackage{pifont}

\begin{document}


\title{General theory for longitudinal nonreciprocal charge transport}

\author{Hong Jian Zhao}
 \affiliation{Key Laboratory of Material Simulation Methods and Software of Ministry of Education, College of Physics, Jilin University, Changchun 130012, China}
\author{Lingling Tao}
 \affiliation{School of Physics, Harbin Institute of Technology, Harbin 150001, China}
\author{Yuhao Fu}
 \affiliation{Key Laboratory of Material Simulation Methods and Software of Ministry of Education, College of Physics, Jilin University, Changchun 130012, China}
 \author{Laurent Bellaiche}
 \affiliation{Physics Department and Institute for Nanoscience and Engineering, University of Arkansas, Fayetteville, Arkansas 72701, USA}
 \author{Yanming Ma}
 \affiliation{Key Laboratory of Material Simulation Methods and Software of Ministry of Education, College of Physics, Jilin University, Changchun 130012, China}
 \affiliation{International Center of Future Science, Jilin University, Changchun 130012, China}
 \affiliation{State Key Laboratory of Superhard Materials, College of Physics, Jilin University, Changchun 130012, China}

\begin{abstract}
The longitudinal nonreciprocal charge transport (NCT) in crystalline materials is a highly nontrivial phenomenon, motivating the design of next generation two-terminal rectification devices (e.g., semiconductor diodes beyond PN junctions). The practical application of such devices is built upon crystalline materials whose longitudinal NCT occurs at room temperature and under low magnetic field. However, materials of this type are rather rare and elusive, and theory guiding the discovery of these materials is lacking. Here, we develop such a theory within the framework of semiclassical Boltzmann transport theory. By symmetry analysis, we classify the complete 122 magnetic point groups with respect to the longitudinal NCT phenomenon. The symmetry-adapted Hamiltonian analysis further uncovers a previously overlooked mechanism for this phenomenon. Our theory guides the first-principles prediction of longitudinal NCT in multiferroic $\varepsilon$-Fe$_2$O$_3$ semiconductor that possibly occurs at room temperature, without the application of external magnetic field. 
These findings advance our fundamental understandings of longitudinal NCT in crystalline materials, and aid the corresponding materials discoveries.
\end{abstract}

\maketitle

\noindent
\textit{Introduction. --} 
The nonreciprocal charge transport (NCT) is a phenomenon for which a material with oppositely flowed electric currents exhibits unequal resistances~\cite{wakatsuki2017,nrp2023,tokura2018}. This phenomenon naturally occurs in semiconductor PN junctions, and yields the two-terminal junction diodes as the building blocks in modern electronics~\cite{diode,nrp2023,tokura2018,np2017}. 
Recent work indicates that crystalline materials with broken inversion and time-reversal symmetries (e.g., noncentrosymmetric semiconductors~\cite{li2021,nctge,np2017,zheliuk2021,nonreciprb2022}, metallic magnets~\cite{liu2022,nimnsb}, and topological materials~\cite{wang2022,zhang2022,wang2023nature}) may host NCT as well~\cite{tokura2018,nonreciprolight}. 
The NCT in crystalline materials is comprised of a transversal part and a longitudinal part, where the latter opens an entirely new route to design novel two-terminal rectification devices (see, e.g., Refs.~\cite{li2021,np2017,ge111,wang2023nature}). 
For instance, the longitudinal NCT in crystalline semiconductors motivates the design of next-generation semiconductor diodes, resembling the diodes based on PN junctions but without involving any junction~\cite{li2021,np2017,ge111,wang2023nature}. Designing such devices and enabling their practical applications rely on crystalline materials with longitudinal NCT at room temperature and under low magnetic field, while this type of materials are rare and elusive. 
To guide materials discovery, a theory capturing the essential physics of longitudinal NCT in crystalline materials is of high necessity. But, unlike the case of PN junctions, the longitudinal NCT phenomenon in crystals is rather complicated~\cite{np2017,tokura2018} --- the aforementioned theory remaining lacking.

Here, we develop a general theory for longitudinal NCT in ferromagnetic, antiferromagnetic, and nonmagnetic crystalline materials, within the framework of Boltzmann transport theory.  
We perform symmetry analysis and provide a classification of the complete 122 magnetic point groups (MPGs) regarding longitudinal NCT. Specifically, we identify 42 MPGs that host {\it intrinsic} longitudinal NCT (without involving magnetic field), where the longitudinal NCT stems from the magnetic order parameter. This resembles the magnetochiral anisotropy effect demonstrated in, for instance, Refs.~\cite{nonrecipro,nonrecipro2,wang2022,tokura2018}. 
We also find 20 MPGs that accommodate the extrinsic longitudinal NCT induced by external magnetic field, namely, the magnetochiral anisotropy effect. 
The longitudinal NCT in crystalline materials is further illustrated by constructing effective Hamiltonians. The effective Hamiltonian analysis helps identify a previously overlooked mechanism responsible for the longitudinal NCT. Motivated by the design of intrinsic semiconductor diodes and guided by our theory, we predict by first-principles simulations that multiferroic $\varepsilon$-Fe$_2$O$_3$ semiconductor showcases intrinsic longitudinal NCT occurring at room temperature. \\

\noindent
\textit{The longitudinal NCT from second-order nonlinear charge current. --} To begin with, we briefly overview the magnetochiral anisotropy effect in crystalline materials (see, e.g., Refs.~\cite{nonrecipro,nonrecipro2,wang2022,tokura2018}). Under external magnetic field $B$, a crystalline material with electric current $I$ gains an unidirectional magnetoresistance $R(B,I)=\xi B I$~\cite{tokura2018,np2017,nonrecipro,nonrecipro2,nagaosa2018prl,nagaosa2018prb}. The sign of $R(B,I)$ is reversed by flipping $I$ or $B$, and this corresponds to the NCT phenomenon. In the following, we shall demonstrate that the longitudinal NCT phenomenon generally occurs in materials with a spontaneous or induced magnetic order parameter $L$ (e.g., magnetization or N\'eel vector), where $L$ plays the role as $B$ in $R(B,I)=\xi B I$. 

We recall that the longitudinal NCT requires the asymmetric band dispersion with respect to wave vector $\mathbf{k}\equiv k_\alpha \boldsymbol{\alpha} + k_\beta \boldsymbol{\beta} + k_\gamma \boldsymbol{\gamma}$ ($\boldsymbol{\alpha}$, $\boldsymbol{\beta}$, and $\boldsymbol{\gamma}$ being three orthogonal unit vectors)~\cite{liu2022,tokura2018,fu2020}. To understand this, a starting point is to examine the longitudinal 
second-order charge current density $J_{\alpha}^{(2)}$ [see e.g., Refs.~\cite{li2021,liu2022,nimnsb,np2017}
and Section I of the Supplementary Material (SM)]. For simplicity, our discussion is limited in the regime of direct current. Within Boltzmann transport theory (constant relaxation time approximation), $J_{\alpha}^{(2)}$ is expressed as~\cite{liu2022}
\begin{eqnarray}\label{eq:chargecurrent2nd}
J_{\alpha}^{(2)}= \frac{e^3\tau^2 E_\alpha^2}{8\pi^3\hbar^3} \sum_n \iiint \frac{\partial^2 \epsilon_n}{\partial k_\alpha^2} \frac{df_0(\epsilon_n)}{d\epsilon_n} \frac{\partial{\epsilon_n}}{\partial{k_\alpha}} \,d^3 \mathbf{k},
\end{eqnarray}
where $\epsilon_n(\mathbf{k})\equiv\epsilon_n$ is the band dispersion, $n$ the band index, $\tau$ the electronic relaxation time, $\hbar$ the reduced Planck constant, $e$ the elementary charge, $f_0(\epsilon_n)$ the Fermi-Dirac distribution at $\epsilon_n$, and $E_\alpha$ the electric field along $\boldsymbol{\alpha}$ direction.

To show that $J_{\alpha}^{(2)}$ arises from the asymmetric band dispersion, we consider a symmetry operation that transforms $\mathbf{k}=k_\alpha \boldsymbol{\alpha} + k_\beta \boldsymbol{\beta} + k_\gamma \boldsymbol{\gamma}$ to $\mathbf{k}^\prime=-k_\alpha \boldsymbol{\alpha} + \tilde{\kappa}_\beta \boldsymbol{\beta} + \tilde{\kappa}_\gamma \boldsymbol{\gamma}$, such that $\epsilon_n(k_\alpha \boldsymbol{\alpha} + k_\beta \boldsymbol{\beta} + k_\gamma \boldsymbol{\gamma})=\epsilon_n(-k_\alpha \boldsymbol{\alpha} + \tilde{\kappa}_\beta \boldsymbol{\beta} + \tilde{\kappa}_\gamma \boldsymbol{\gamma})$ --- the band dispersion $\epsilon_n(\mathbf{k})$ being symmetric with respect to $k_\alpha$. This implies that $\partial{\epsilon_n(\mathbf{k})}/\partial{k_\alpha}$ at $\mathbf{k}$ and $\mathbf{k}^\prime$ are opposite numbers, while the other two quantities [i.e., $\partial^2 \epsilon_n(\mathbf{k})/\partial k_\alpha^2$ and $df_0(\epsilon_n)/d\epsilon_n$] are identical. Associated with each $\epsilon_n$, the integral function in Eq.~(\ref{eq:chargecurrent2nd}) cancels out over the integration region, and this yields null $J_{\alpha}^{(2)}$. To achieve non-zero $J_{\alpha}^{(2)}$, the linkage between $k_\alpha$ and $-k_\alpha$ must be broken, namely, $ \epsilon_n(k_\alpha \boldsymbol{\alpha} + k_\beta \boldsymbol{\beta} + k_\gamma \boldsymbol{\gamma})$ is never symmetrically related to $\epsilon_n(-k_\alpha \boldsymbol{\alpha} + \tilde{\kappa}_\beta \boldsymbol{\beta} + \tilde{\kappa}_\gamma \boldsymbol{\gamma})$ no matter what $\tilde{\kappa}_\beta$ and $\tilde{\kappa}_\gamma$ are selected. In view of this, $J_{\alpha}^{(2)}$ only occurs in materials with specific symmetry constraints. For example, materials with time-reversal symmetry $\mathfrak{1}^\prime$ do not host $J_{\alpha}^{(2)}$, because $\mathfrak{1}^\prime$ link $k_\alpha \boldsymbol{\alpha} + k_\beta \boldsymbol{\beta} + k_\gamma \boldsymbol{\gamma}$ with $-k_\alpha \boldsymbol{\alpha} - k_\beta \boldsymbol{\beta} - k_\gamma \boldsymbol{\gamma}$. On the contrary, materials with magnetic order parameter $L$ (i.e., broken time-reversal symmetry) might be compatible with $J_{\alpha}^{(2)}$~\footnote{The work ``might'' means that the broken time-reversal symmetry is not the sufficient condition for the nonreciprocal charge transport. As a matter of fact, the other symmetry operations (e.g., spatial inversion) also constrain the behaviors of nonreciprocal charge transport. This will be demonstrated in the following section.}. As analyzed in Section I of the SM, $J_{\alpha}^{(2)}$ is a function of $L$, and the nonlinear longitudinal conductivity $\sigma_{\alpha\alpha\alpha}^{(2)}$ is given by 
\begin{eqnarray}\label{eq:chargecurrent2nd2}
\sigma_{\alpha\alpha\alpha}^{(2)} = \frac{\zeta(L)e^3\tau^2}{8\pi^3\hbar^3} \sum_n \iiint \frac{\partial^2 \epsilon_n}{\partial k_\alpha^2} \frac{df_0(\epsilon_n)}{d\epsilon_n} \frac{\partial{\epsilon_n}}{\partial{k_\alpha}} \,d^3 \mathbf{k},
\end{eqnarray}
where $\zeta(L)=\pm 1$ and $\zeta(-L)=-\zeta(L)$ indicate the dependence of $\sigma_{\alpha\alpha\alpha}^{(2)}$ on $L$~\footnote{In materials with magnetic order parameter $L$, the first-order linear charge current remains invariant when reversing $L$ (e.g., from $L$ to $-L$). The detailed analysis can be found in Section I of the SM.}. We now show that $\sigma_{\alpha\alpha\alpha}^{(2)}$ contributes to longitudinal NCT. To this end, we consider the total longitudinal charge current density $J_\alpha=\sigma^{(1)}_{\alpha\alpha}E_\alpha+\sigma_{\alpha\alpha\alpha}^{(2)} E_\alpha^2$~\cite{nimnsb,nagaosa2018prb}, with $\sigma^{(1)}_{\alpha\alpha}$ being the linear conductivity; In first approximation, the electric field is expressed as $E_\alpha\approx J_\alpha/\sigma^{(1)}_{\alpha\alpha}$. 
This suggests an effective conductivity $\sigma^\mathrm{eff}_{\alpha\alpha}=\sigma^{(1)}_{\alpha\alpha}+2\sigma_{\alpha\alpha\alpha}^{(2)}E_\alpha \approx  \sigma^{(1)}_{\alpha\alpha}+2\sigma_{\alpha\alpha\alpha}^{(2)} J_\alpha/\sigma^{(1)}_{\alpha\alpha} \equiv \sigma^{(1)}_{\alpha\alpha} + \xi_\alpha J_\alpha \zeta(L)$ ($\xi_\alpha$ being a coefficient). The term $\xi_\alpha J_\alpha \zeta(L)$ resembles $R(B)=\xi BI$ as follows: reversing $L$ or $J_\alpha$ changes the sign of $\sigma_{\alpha\alpha\alpha}^{(2)}$, where $L$ and $J_\alpha$ play the roles as $B$ and $I$, respectively. In other words, the nonlinear conductivity $\sigma_{\alpha\alpha\alpha}^{(2)}$ characterizes the longitudinal NCT along $\alpha$ direction. \\

\begin{table}[ht]
\caption{\label{tab:pointmag} The 42 MPGs that allow the longitudinal NCT. For each MPG, the \ding{51} and \ding{55} indicate that longitudinal NCT along $\alpha$ direction is symmetrically allowed and forbidden, respectively. Here, $\alpha=x,y,z$ marks the direction in Cartesian frame. The conventions regarding the coordinate system for these MPGs are shown in Table S1 of the SM.} 
\begin{ruledtabular}
\begin{tabular}{lccc|lccc|lccc}
MPGs & $x$ & $y$ & $z$ & MPGs & $x$ & $y$ & $z$ & MPGs & $x$ & $y$ & $z$ \\
 \hline
$1.1$ & \ding{51} & \ding{51} & \ding{51} &
 $\bar{1}^\prime$ & \ding{51} & \ding{51} & \ding{51}  & 
$2.1$ & \ding{55} & \ding{55} & \ding{51} \\
$2^\prime$ & \ding{51} & \ding{51} & \ding{55} &
$m.1$  & \ding{51} & \ding{51} & \ding{55} &
$m^\prime$ & \ding{55} & \ding{55} & \ding{51} \\
$2^\prime/m$ & \ding{51} & \ding{51} & \ding{55} &
$2/m^\prime$ & \ding{55} & \ding{55} & \ding{51} &
$2^\prime2^\prime2$ & \ding{55} & \ding{55} & \ding{51} \\
$mm2.1$ & \ding{55} & \ding{55} & \ding{51} &
$m^\prime m 2^\prime$ & \ding{51} & \ding{55} & \ding{55} &
$m^\prime m m$ & \ding{51} & \ding{55} & \ding{55} \\
$4.1$ & \ding{55} & \ding{55} & \ding{51} &
$\bar{4}^\prime$ & \ding{55} & \ding{55} & \ding{51} &
$4/m^\prime$ & \ding{55} & \ding{55} & \ding{51} \\
$4 2^\prime 2^\prime$  & \ding{55} & \ding{55} & \ding{51} &
$4mm.1$ & \ding{55} & \ding{55} & \ding{51} &
$\bar{4}^\prime 2^\prime m $ & \ding{55} & \ding{55} & \ding{51} \\
$4/m^\prime mm$ & \ding{55} & \ding{55} & \ding{51} &
$3.1$ & \ding{51} & \ding{51} & \ding{51} &
$\bar{3}^\prime$ & \ding{51} & \ding{51} & \ding{51} \\
$32.1$ & \ding{51} & \ding{55} & \ding{55} &
$32^\prime$ & \ding{55} & \ding{51} & \ding{51} &
$3m.1$  & \ding{55} & \ding{51} & \ding{51} \\
$3m^\prime$ & \ding{51} & \ding{55} & \ding{55} &
$\bar{3}^\prime m$ & \ding{55} & \ding{51} & \ding{51} &
$\bar{3}^\prime m^\prime$ & \ding{51} & \ding{55} & \ding{55} \\
$6.1$ & \ding{55} & \ding{55} & \ding{51} &
$6^\prime$ & \ding{51} & \ding{51} & \ding{55} &
$\bar{6}.1$ & \ding{51} & \ding{51} & \ding{55} \\
$\bar{6}^\prime$ & \ding{55} & \ding{55} & \ding{51} &
$6^\prime/m$ & \ding{51} & \ding{51} & \ding{55} &
$6/m^\prime$ & \ding{55} & \ding{55} & \ding{51} \\
$6^\prime 2 2^\prime$ & \ding{51} & \ding{55} & \ding{55} &
$6 2^\prime 2^\prime$ & \ding{55} & \ding{55} & \ding{51} &
$6mm.1$ & \ding{55} & \ding{55} & \ding{51} \\
$6^\prime m m^\prime$ & \ding{55} & \ding{51} & \ding{55} &
$\bar{6}m2.1$ & \ding{55} & \ding{51} & \ding{55} &
$\bar{6}^\prime m 2^\prime$ & \ding{55} & \ding{55} & \ding{51} \\
$\bar{6}m^\prime 2^\prime$ & \ding{51} & \ding{55} & \ding{55} &
$6/m^\prime mm$  & \ding{55} & \ding{55} & \ding{51} &
$6^\prime/mmm^\prime$ & \ding{55} & \ding{51} & \ding{55} \\  
\end{tabular}
\end{ruledtabular}
\end{table}

\noindent
\textit{Symmetry analysis. --} We move on to carry out symmetry analysis regarding the longitudinal NCT. We use the $m^\prime m 2^\prime$ magnetic point group (MPG) to demonstrate our basic ideas. This MPG contains four symmetry operations, namely, $\mathfrak{1}$, $\mathfrak{m}_y$, $\mathfrak{m}_x^\prime$, and $\mathfrak{2}_z^\prime$. The $\mathfrak{1}$ symmetry operation is the identity, and has no effect on $(k_x,k_y,k_z)\equiv k_x \mathbf{x}+ k_y \mathbf{y} + k_z \mathbf{z}$ ($\mathbf{x},\mathbf{y},\mathbf{z}$ being unit vectors along the Cartesian $x,y,z$ directions). The $\mathfrak{m}_y$ operation is a mirror plane perpendicular to $y$, and it transforms $(k_x,k_y,k_z)$ to $(k_x,-k_y,k_z)$. The $\mathfrak{m}_x^\prime$ operation, the mirror plane perpendicular to $x$ followed by a time-reversal operation, transforms $(k_x,k_y,k_z)$ to $(k_x,-k_y,-k_z)$. Finally, $(k_x,k_y,k_z)$ is transformed to $(k_x,k_y,-k_z)$ by $\mathfrak{2}_z^\prime$, the twofold rotation along $z$ followed by a time-reversal. On balance, the symmetry operations of the $m^\prime m 2^\prime$ MPG (i) link $k_y$ with $-k_y$ by $\mathfrak{m}_y$ or $\mathfrak{m}_x^\prime$, (ii) link $k_z$ with $-k_z$ by $\mathfrak{m}_x^\prime$ or $\mathfrak{2}_z^\prime$, and (iii) provide no linkage between $k_x$ and $-k_x$. This means that the longitudinal NCT in $m^\prime m 2^\prime$ MPG is symmetrically allowed along $x$ direction.

In this way, we conduct symmetry analysis on the complete 122 MPGs. These groups are composed of 32 type-1 MPGs, 32 type-2 MPGs, and 58 type-3 MPGs, where type-2 MPGs contain time-reversal symmetry $\mathfrak{1}^\prime$, but type-1 and type-3 MPGs do not have $\mathfrak{1}^\prime$~\cite{uni2022,symmetry}. The difference between type-1 and type-3 MPGs is that the former contain only the spatial symmetry operations, while the latter also involve some symmetry operations being a spatial operation followed by $\mathfrak{1}^\prime$ (see e.g., Ref.~\cite{uni2022} for details). Among the type-1 and type-3 MPGs, 42 cases host the symmetrically allowed longitudinal NCT (see Table~\ref{tab:pointmag}). 
Next, we consider 32 type-2 MPGs containing time-reversal symmetry. These type-2 MPGs forbid the longitudinal NCT; Nonetheless, magnetic field breaks time-reversal and other symmetries in type-2 MPGs, possibly yielding longitudinal NCT. Regarding this, we analyze the magnetic field induced symmetry breakings in type-2 MPGs, and identify 20 cases in which magnetic field enables longitudinal NCT (see Table~\ref{tab:pointgray}). The detailed analysis is shown in Section II of the SM.\\

\begin{table}[ht]
\caption{\label{tab:pointgray} The magnetic field induced longitudinal NCT in 20 type-2 MPGs. The $B_x$, $B_y$, and $B_z$ marks the $x$, $y$, and $z$ components of the magnetic field, respectively. The directions for longitudinal NCT are labelled by $x$, $y$, and $z$ (Cartesian frame). The conventions regard the coordinate system for these MPGs are shown in Table S1 of the SM.} 
\begin{ruledtabular}
\begin{tabular}{lccc|lccc}
MPGs & $B_x$ & $B_y$ & $B_z$ & MPGs & $B_x$ & $B_y$ & $B_z$ \\
 \hline
$1.1^\prime$ & $x,y,z$ & $x,y,z$ & $x,y,z$ & $2.1^\prime$ & $x,y$ & $x,y$ & $z$ \\
$m.1^\prime$ & $z$ & $z$ & $x,y$ & $222.1^\prime$ & $x$ & $y$ & $z$ \\
$mm2.1^\prime$ & $y$ & $x$ & --- & $4.1^\prime$ & $x,y$ & $x,y$ & $z$ \\
$\bar{4}.1^\prime$ & $x,y$ & $x,y$ & --- & $422.1^\prime$ & $x$ & $y$ & $z$ \\
$4mm.1^\prime$ & $y$ & $x$ & --- & $\bar{4}2m.1^\prime$ & $x$ & $y$ & --- \\
$3.1^\prime$ & $x,y,z$ & $x,y,z$ & $x,y,z$ & $32.1^\prime$ & $x$ & $y,z$ & $y,z$ \\
$3m.1^\prime$ & $y,z$ & $x$ & $x$ & $6.1^\prime$ & $x,y$ & $x,y$ & $z$ \\
$\bar{6}.1^\prime$ & $z$ & $z$ & $x,y$ & $622.1^\prime$ & $x$ & $y$ & $z$ \\
$6mm.1^\prime$ & $y$ & $x$ & --- &  $\bar{6}m2.1^\prime$ & $z$ & --- &  $x$  \\
$23.1^\prime$ & $x$ & $y$ & $z$ & $432.1^\prime$ & $x$ & $y$ & $z$ \\
\end{tabular}
\end{ruledtabular}
\end{table}

\noindent
\textit{Effective Hamiltonians for longitudinal NCT. --} In this section, we explore the role of magnetic order parameter $L$ in band asymmetry and longitudinal NCT. For this purpose, we derive the minimal two-band effective Hamiltonians for the 42 MPGs listed in Table~\ref{tab:pointmag}, involving the magnetic order parameter $L$, the wave vector $\mathbf{k}$, and the electronic spin $\boldsymbol{\sigma}\equiv(\sigma_x,\sigma_y,\sigma_z)$ --- $\boldsymbol{\sigma}$ being Pauli matrix vector. The results are summarized in Table S3 and Table S4 of the SM. The effective Hamiltonians (around the center of the Brillouin zone) for these 42 MPGs are generally written as
\begin{eqnarray}\label{eq:heffnct}
H(\mathbf{k},L)  = \sum_{\alpha,\beta=x,y,z} \mu_{\alpha\beta}k_\alpha k_\beta \sigma_0 + \zeta(L) \Lambda(\mathbf{k}) \sigma_0 \nonumber \\
 + \boldsymbol{\lambda}(\mathbf{k})\cdot\boldsymbol{\sigma}  + \zeta(L) \boldsymbol{\Delta} \cdot\boldsymbol{\sigma},
\end{eqnarray}
where $\mu_{\alpha\beta}$, $\Lambda(\mathbf{k})$, $\boldsymbol{\lambda}(\mathbf{k})$, and $\boldsymbol{\Delta}$ characterize the effective mass, band asymmetry, spin-orbit field, and Zeeman field, respectively ($\sigma_0$ being $2\times2$ identity matrix). The effective mass terms and band asymmetry terms appear in the effective Hamiltonians of all these 42 MPGs. Furthermore, MPGs lacking the parity-time symmetry (i.e., inversion followed by time-reversal) may also have spin-orbit field terms, while MPGs compatible with ferromagnetism extra gain Zeeman field terms. 
Of particular interest is the $\zeta(L) \Lambda(\mathbf{k}) \sigma_0$ band asymmetry term, with $\Lambda(\mathbf{k})$ being an odd function of $k_\chi$. 
Such a term describes the band asymmetry with respect to $k_\chi$ as well as the longitudinal NCT along $\chi$ direction. As for the spin-orbit field and Zeeman field terms, the situation becomes quite complicated. This will be discussed in the following paragraphs.

\begin{figure}[t!]
\centering
\includegraphics[width=1\linewidth]{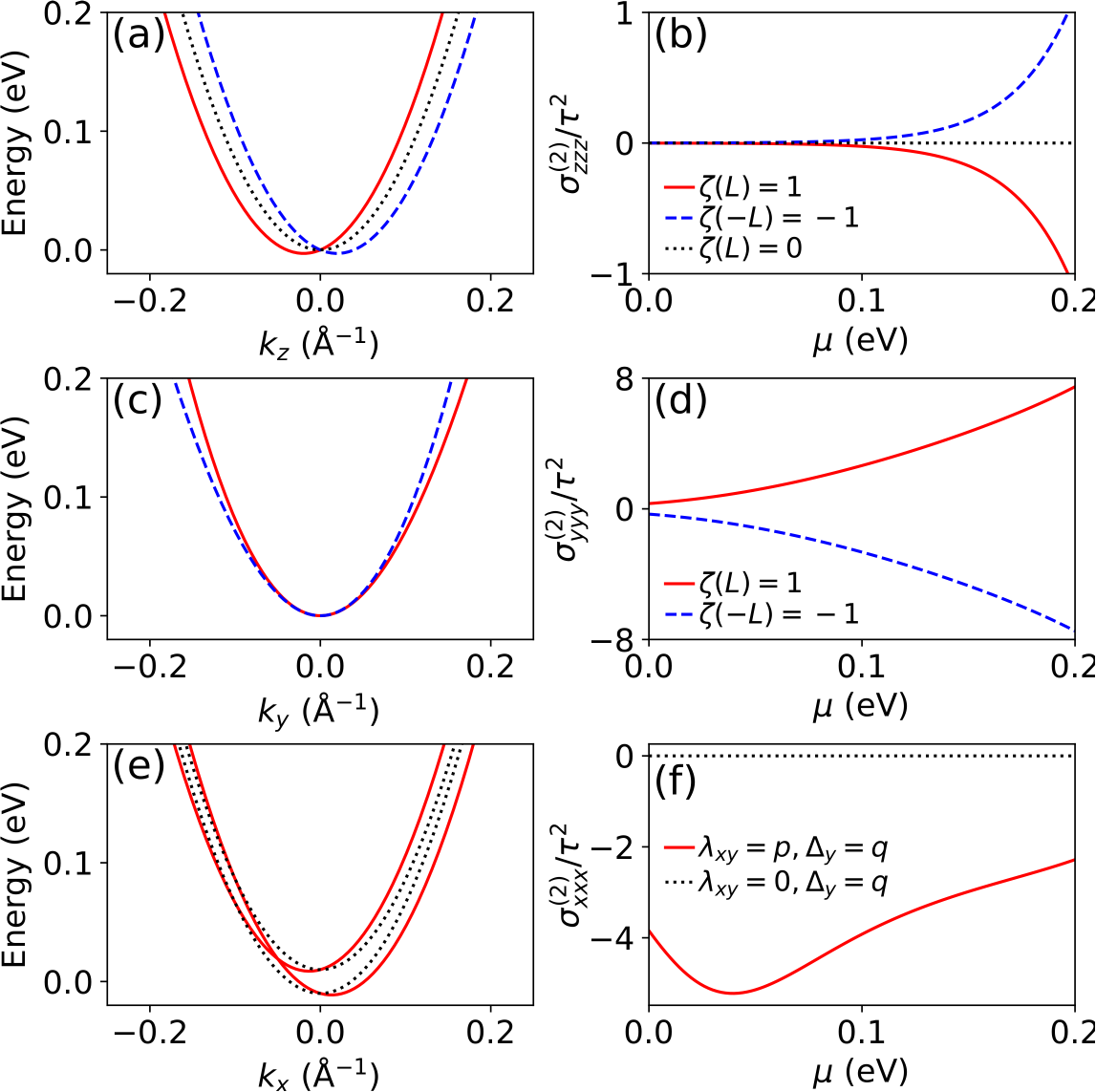}
\caption{\label{fig:toymodel} The band structures and longitudinal NCT obtained from different effective Hamiltonians. (a) and (b): $H_2(\mathbf{k},L)  = \mu_{xx} (k_x^2 + k_y^2)\sigma_0 + \mu_{zz} k_z^2 \sigma_0 + \zeta(L)\Lambda_z k_z \sigma_0$ with $\Lambda_z=0.3$ eV {\AA}.  (c) and (d): $H_3(\mathbf{k},L) = \mu_{xx}(k_x^2+k_y^2) \sigma_0 + \mu_{zz}k_z^2 \sigma_0 + \zeta(L) \Lambda_{yyy} k_y (3 k_x^2 - k_y^2) \sigma_0$ with $\Lambda_{yyy}=5.0$ eV {\AA}$^3$. (e) and (f): $H_4(\mathbf{k},L)  = (\mu_{xx} k_x^2 + \mu_{yy} k_y^2 + \mu_{zz} k_z^2)\sigma_0 + \zeta(L)\Lambda_x k_x \sigma_0 + \zeta(L) \Delta_y \sigma_y  + \lambda_{xy} k_x\sigma_y + \lambda_{yx} k_y\sigma_x$, with $\Lambda_x=0.0$ eV {\AA}, $\lambda_{yx}=0.3$ eV {\AA}, $p=0.2$ eV {\AA}, $q=0.01$ eV, and $\zeta(L)= 1$. Here, $\zeta(L)=1$ and $\zeta(-L)=-1$ corresponds to $L$ and $-L$ magnetic order parameters, respectively. As for $H_2(\mathbf{k},L)$, $H_3(\mathbf{k},L)$, and $H_4(\mathbf{k},L)$, the coefficients $\mu_{xx}$, $\mu_{yy}$, and $\mu_{zz}$ are set as $\mu_{xx}=\mu_{yy}=\mu_{zz}=\hbar^2/2m=7.62$ eV {\AA}$^2$, where $m=0.5 m_0$ and $m_0$ is electron rest mass. The unit of $\sigma^{(2)}_{zzz}/\tau^2$, $\sigma^{(2)}_{yyy}/\tau^2$, and $\sigma^{(2)}_{xxx}/\tau^2$ is $10^{23}~\Omega^{-1}\mathrm{V}^{-1}\mathrm{s}^{-2}$. The thermal smearing with a temperature of 300 K is adopted during the conductivity calculations. The legends for (a), (c), and (d) are shown in (b), (d), and (f), respectively.}
\end{figure}

We now take a few representative MPGs to perform our Hamiltonian analysis. 
Our first example is the $6mm.1$ MPG with its effective Hamiltonian given by $H_1(\mathbf{k},L)=\mu_{xx}(k_x^2+k_y^2) \sigma_0 + \mu_{zz}k_z^2 \sigma_0 + \zeta(L) \Lambda_{z} k_z \sigma_0 + \lambda_{xy} (k_x \sigma_y - k_y \sigma_x)$. This Hamiltonian contains the effective mass terms, the band asymmetry term, and the spin-orbit field terms (no Zeeman field terms). 
Some other MPGs may have effective Hamiltonians with only effective mass terms and band asymmetry terms. For instance, the effective Hamiltonians for 
$4/m^\prime mm$ and $6^\prime/m m m^\prime$ MPGs are $H_2(\mathbf{k},L)  = \mu_{xx} (k_x^2 + k_y^2)\sigma_0 + \mu_{zz} k_z^2 \sigma_0 + \zeta(L)\Lambda_z k_z \sigma_0$ and $H_3(\mathbf{k},L) = \mu_{xx}(k_x^2+k_y^2) \sigma_0 + \mu_{zz}k_z^2 \sigma_0 + \zeta(L) \Lambda_{yyy} k_y (3 k_x^2 - k_y^2) \sigma_0$, respectively.
The role of $L$ on the 
band asymmetry and longitudinal NCT can be illustrated by numerically solving $H_2(\mathbf{k},L)$ and $H_3(\mathbf{k},L)$, with various groups
of selected model parameters. As shown in Fig.~\ref{fig:toymodel}(a), the non-zero $\zeta(L) \Lambda_{z}$ results in band asymmetry along $k_z$ direction, where the $-L$ and $+L$ magnetic order parameters yield two version of bands (red and blue lines) being mirror copies
of each other with respect to $k_z=0$. This is responsible for the longitudinal nonreciprocal $\sigma^{(2)}_{zzz}$ electric conductivity,
whose sign is reversed by switching magnetic order parameters between $L$ and $-L$ [Fig.~\ref{fig:toymodel}(b)]. When removing the magnetic order parameter $L$ [i.e., $\zeta(L)=0$], both the band asymmetry and longitudinal $\sigma^{(2)}_{zzz}$ conductivity vanish [see Figs.~\ref{fig:toymodel}(a) and~\ref{fig:toymodel}(b)]. As for $H_3(\mathbf{k},L)$, the $\zeta(L) \Lambda_{yyy} k_y (3 k_x^2 - k_y^2) \sigma_0$ term is cubic in $k_y$, which yields the band asymmetry and longitudinal NCT along $y$ [see Figs.~\ref{fig:toymodel}(c) and~\ref{fig:toymodel}(d)]. Various MPGs (e.g., $\bar{1}^\prime$, $m^\prime m m$, and $\bar{3}^\prime$) have effective Hamiltonians similar to $H_1(\mathbf{k},L)$, $H_2(\mathbf{k},L)$, or $H_3(\mathbf{k},L)$, that is, with band asymmetry terms and without Zeeman field terms. In such Hamiltonians, the longitudinal NCT is solely governed by the band asymmetry terms, which is spin independent.

The $m^\prime m 2^\prime$ is another exemplified MPG with an effective Hamiltonian $H_4(\mathbf{k},L)  = (\mu_{xx} k_x^2 + \mu_{yy} k_y^2 + \mu_{zz} k_z^2)\sigma_0 + \zeta(L)\Lambda_x k_x \sigma_0 + \zeta(L) \Delta_y \sigma_y  + \lambda_{xy} k_x\sigma_y + \lambda_{yx} k_y\sigma_x$. Such a Hamiltonian contains the effective mass terms, the spin-orbit field terms, a band asymmetry term, and a Zeeman field term. Regarding $H_4(\mathbf{k},L)$, there are two mechanisms responsible for the longitudinal NCT. First of all, the $\zeta(L)\Lambda_x k_x \sigma_0$ term suggests a longitudinal NCT along $x$ direction. This mechanism has already been discussed in the last paragraph. The second mechanism comes from the combination of spin-orbit field term $\lambda_{xy} k_x\sigma_y$ 
and Zeeman field term $\zeta(L) \Delta_y \sigma_y$, which gives rise to band asymmetry along $k_x$ and longitudinal $\sigma^{(2)}_{xxx}$ conductivity [see Figs.~\ref{fig:toymodel}(e) and~\ref{fig:toymodel}(f)]. This situation likely occurs when spin-orbit field and Zeeman field cooperatively breaks the symmetric linkage between $k_x$ and $-k_x$. Without $\lambda_{xy} k_x\sigma_y$ or $\zeta(L)\Lambda_x k_x \sigma_0$, the $\zeta(L) \Delta_y \sigma_y$ term can not solely generate band asymmetry or longitudinal NCT [see Figs.~\ref{fig:toymodel}(e) and~\ref{fig:toymodel}(f)]. Previous studies usually consider the spin-orbit field terms and the Zeeman field terms, but neglecting the $\zeta(L)\Lambda(\mathbf{k})\sigma_0$ term (see e.g., Refs.~\cite{zhang2022,np2017,nagaosa2018prb,nonreciprb2022,tao2020}). Even though the combination of $\zeta(L)\Delta_\alpha \sigma_\alpha$ and 
$\lambda_\alpha(\mathbf{k})\sigma_\alpha$ might capture the longitudinal NCT, there are no reasons to ignore the $\zeta(L)\Lambda(\mathbf{k})\sigma_0$ term. As a matter of fact, materials with negligible $\Delta_\alpha$ and/or $\lambda_\alpha(\mathbf{k})$ may still have sizable $\Lambda(\mathbf{k})$, and in that case, neglecting $\zeta(L)\Lambda(\mathbf{k}) \sigma_0$ will erroneously describe the longitudinal NCT. \\

\begin{figure}[t!]
\centering
\includegraphics[width=1\linewidth]{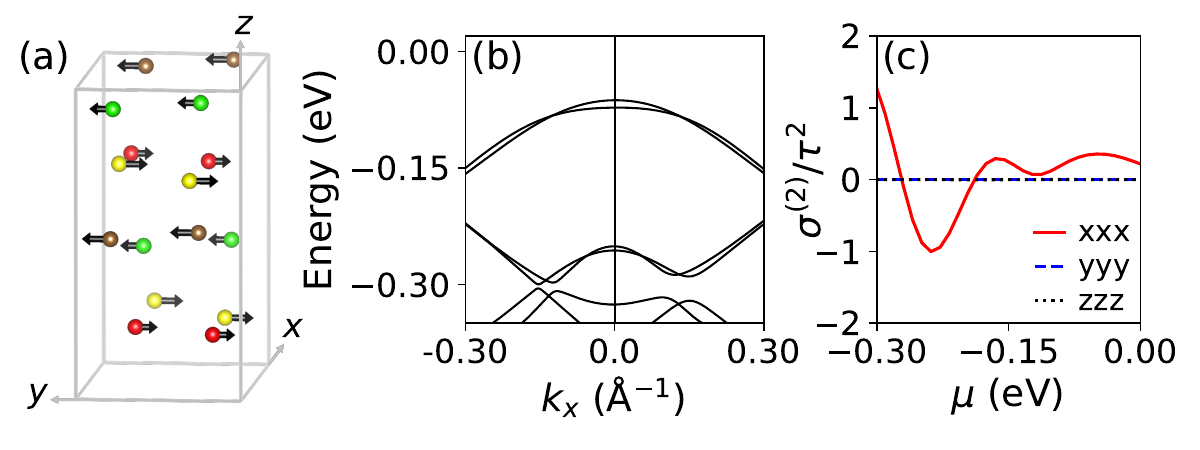}
\caption{\label{fig:fe2o3} The magnetic order parameter (a), band dispersion (b), and longitudinal NCT conductivity (c) of $\varepsilon$-Fe$_2$O$_3$. In (a), the brown, yellow, red, and green spheres denote Fe1, Fe2, Fe3 and Fe4 sublattices (O ions not shown), respectively. The arrows represent Fe's magnetic moments in Fe1 ($3.6~\mu_B$), Fe2 ($-3.6~\mu_B$), Fe3 ($-2.5~\mu_B$), and Fe4 ($2.7~\mu_B$) sublattices~\cite{fe2o3}; Only the predominant components of these magnetic moments are shown. In (b), the band dispersion is along $k_x$ direction, where $k_y$ and $k_z$ are set to zero. The valence band maximum is set as the zero energy. In (c), $\mu$ denote the chemical potential. The red solid, blue dash, and black dot lines represent $\sigma^{(2)}_{xxx}/\tau^2$, $\sigma^{(2)}_{yyy}/\tau^2$ and $\sigma^{(2)}_{zzz}/\tau^2$, respectively. The unit of $\sigma^{(2)}/\tau^2$ is $10^{23}~\Omega^{-1}\mathrm{V}^{-1}\mathrm{s}^{-2}$.}
\end{figure}

\noindent
\textit{The longitudinal NCT in $\varepsilon$-Fe$_2$O$_3$. --} Our Tables~\ref{tab:pointmag} and ~\ref{tab:pointgray} guide the discovery of materials with longitudinal NCT. We are motivated by the design of intrinsic semiconductor diodes, and decide to seek semiconductors with longitudinal NCT. Searching from the MAGNDATA database~\cite{magndata}, we identify $\varepsilon$-Fe$_2$O$_3$ as a promising candidate material. $\varepsilon$-Fe$_2$O$_3$ is a multiferroic semiconductor, being the metastable phase of Fe$_2$O$_3$~\cite{fe2o3,fe2o32,fe2o33,xu2018}. Recently, single crystals of $\varepsilon$-Fe$_2$O$_3$ were experimentally synthesized~\cite{fe2o33}. At room temperature, $\varepsilon$-Fe$_2$O$_3$ has the polar $m^\prime m 2^\prime$ MPG, with the magnetic order parameter schematized in Fig.~\ref{fig:fe2o3}(a)~\cite{fe2o3,fe2o32,magndata}. As shown in Table~\ref{tab:pointmag}, the longitudinal NCT along the $x$ direction is symmetrically allowed in $m^\prime m 2^\prime$ MPG.

To validate the longitudinal NCT in $\varepsilon$-Fe$_2$O$_3$, we conduct first-principles simulations and electric conductivity calculations. We recall that temperature complicatedly affects the conductivity of magnetic materials~\cite{ssp} via, for instance, (i) modifying the magnetic order parameter and the resultant band dispersions, and (ii) changing the Fermi-Dirac distribution and the resultant band occupancy. Therefore, the accurate calculations of conductivity at finite temperatures are rather challenging. As for $\varepsilon$-Fe$_2$O$_3$, we are inclined to qualitatively examine the possibility towards the longitudinal NCT at room temperature, as opposed to quantitatively determining its electric conductivity values.

The Fe's magnetic moments in Fe1, Fe2, Fe3, and Fe4 sublattices of $\varepsilon$-Fe$_2$O$_3$ are $3.9$, $-3.9$, $-4.0$, and $3.9$ $\mu_B$, respectively, as obtained by first-principles simulations. This basically coincides with the experimental results in Fig.~\ref{fig:fe2o3}(a), where the difference comes from the fact that first-principles calculations neglect temperature effects on the magnetic order parameter. Subsequently, we calculate the band dispersion and nonlinear conductivity of multiferroic $\varepsilon$-Fe$_2$O$_3$~\footnote{We neglect the temperature effects on the magnetic order parameter and band dispersion of $\varepsilon$-Fe$_2$O$_3$.}. Figure~\ref{fig:fe2o3}(b) shows the band structure of $\varepsilon$-Fe$_2$O$_3$ as a function of $k_x$. The band dispersion along $k_x$ is significantly asymmetric in the energy range of $\sim-0.3$ eV to $\sim-0.2$ eV. As shown in Fig.~\ref{fig:fe2o3}(c), the nonlinear conductivity values $\sigma^{(2)}_{yyy}$ and $\sigma^{(2)}_{zzz}$ are negligible, while $\sigma^{(2)}_{xxx}$ is finite. This verifies our aforementioned symmetry arguments on $\varepsilon$-Fe$_2$O$_3$. Our calculations, although based on the ground state of $\varepsilon$-Fe$_2$O$_3$, correctly reflect the MPG of such a material at room temperature. This suggests that $\varepsilon$-Fe$_2$O$_3$ may host room-temperature longitudinal NCT that is driven by its intrinsic magnetic order parameter (i.e., without the application of external magnetic field).\\

\noindent
\textit{Summary and perspective. --} In summary, we have developed a general theory guiding the discovery of crystalline materials with longitudinal NCT. Within the framework of Boltzmann transport theory, the longitudinal NCT in crystalline materials resides in the band asymmetry, and is reflected by the second-order nonlinear longitudinal conductivity. Based on this, we provide a comprehensive symmetry classification of the 122 MPGs with respect to longitudinal NCT (see Tables~\ref{tab:pointmag} and~\ref{tab:pointgray}). By constructing and analyzing effective Hamiltonians, we identify two mechanisms for longitudinal NCT, that is, the band asymmetry $\Lambda(\mathbf{k})$, and the combination of spin-orbit field $\boldsymbol{\lambda}(\mathbf{k})$ and Zeeman field $\boldsymbol{\Delta}$ [see Eq.~(\ref{eq:heffnct})]. Our theory together with first-principles simulations help to identify multiferroic $\varepsilon$-Fe$_2$O$_3$ as a candidate that possibly showcases intrinsic longitudinal NCT at room temperature. 

Beyond this, our theory also suggests another research avenue. As shown in Fig.~\ref{fig:toymodel}, the longitudinal NCT severely depends on the magnetic order parameters. For a specific material with MPG listed in Tables~\ref{tab:pointmag} and~\ref{tab:pointgray}, the measurement of nonlinear longitudinal conductivity reflects its intrinsic magnetic ordering or the external magnetic field applied to it. In this regard, the longitudinal NCT together with second-order transverse Drude transport and second-order anomalous Hall effect (i.e., second-order nonlinear transport) open a door for the electrical detection of magnetic states~\cite{liu2022,symminhe,symminhe2}, being important for designing spintronic devices~\cite{magdet,afm1,afm4,godinho2018}. Interested readers are referred to Refs.~\cite{symminhe,symminhe2,nsrzhang2023} for some detailed discussion on second-order nonlinear transport.
To finish, we hope that our theory can not only provide in-depth insights into the NCT phenomena in condensed matter, but also guide the materials discovery and device design related to such a phenomenon.\\

\noindent
\textit{Acknowledgements. --}  We acknowledge the support from the National Natural Science Foundation of China
(Grants Nos. 12274174, 12274102, 22090044, 52288102, and 12034009), the National Key Research and Development Program of China (Grant No. 2022YFA1402501), and the Fundamental Research Funds for the Central Universities (Grant No. FRFCU5710053421, No. HIT.OCEF.2023031). L.B. thanks  the Vannevar Bush Faculty Fellowship (VBFF) grant No. N00014-20-1-2834 from the Department of Defense and award No. DMR-1906383 from the National Science Foundation AMASE-i Program (MonArk NSF Quantum Foundry).

%

\end{document}